\documentclass[sigconf]{acmart}

\copyrightyear{2020}
\acmYear{2020}
\setcopyright{iw3c2w3}
\acmConference[WWW '20]{Proceedings of The Web Conference 2020}{April 20--24, 2020}{Taipei, China}
\acmBooktitle{Proceedings of The Web Conference 2020 (WWW '20), April 20--24, 2020, Taipei, China}
\acmPrice{}
\acmDOI{10.1145/3366423.3380291}
\acmISBN{978-1-4503-7023-3/20/04}

\usepackage{verbatim}
\usepackage{amsfonts}
\usepackage{algorithm, algorithmicx, algpseudocode}
\usepackage{amsmath}
\usepackage{array}
\usepackage{subcaption}
\usepackage{balance}
\usepackage{multicol}
\usepackage{multirow}
\usepackage{color}
\usepackage{epsfig}
\usepackage{natbib}
\usepackage{xspace}
\usepackage{booktabs}
\usepackage{bm}
\usepackage{enumitem}
\usepackage{diagbox}
\usepackage{url}
\usepackage{geometry}
\usepackage{changepage}
\usepackage{textcomp}
\usepackage[bottom]{footmisc}
\newcommand{\vpara}[1]{\vspace{0.3em}\noindent\textbf{#1 }}
\newcommand{\equationref}[1]{Eq~\ref{#1}}
\newcommand{\secref}[1]{Section~\ref{#1}}

\newcommand{\figref}[1]{Figure~\ref{#1}}
\newcommand{\tableref}[1]{Table~\ref{#1}} 

\newcommand{\methodname}{Hierarchical Electricity-theft Behavior Recognition\xspace}
\newcommand{\methodshort}{HEBR\xspace}

\begin{document}

\title{Understanding Electricity-Theft Behavior via Multi-Source Data}

\author{Wenjie Hu}
\affiliation{
	\institution{Zhejiang University}
}
\email{aston2une@zju.edu.cn }

\author{Yang Yang$^*$}\thanks{$^*$Corresponding author: Yang Yang, \url{yangya@zju.edu.cn} \\$^1$The source code is published at \url{https://github.com/zjunet/HEBR}}
\affiliation{
	\institution{Zhejiang University}
}
\email{yangya@zju.edu.cn}

\author{Jianbo Wang}
\affiliation{
	\institution{State Grid %Zhejiang Taizhou 
		Power Supply Co. Ltd.}
}
\email{jianbosgcc_lq@163.com}

\author{Xuanwen Huang}
\affiliation{
	\institution{Zhejiang University}
}
\email{xwhuang@zju.edu.cn}

\author{Ziqiang Cheng}
\affiliation{
	\institution{Zhejiang University}
}
\email{petecheng@zju.edu.cn}

\renewcommand{\shortauthors}{Hu et al.}

%\begin{CCSXML}
%	<ccs2012>
%	<concept>
%	<concept_id>10010405.10010455</concept_id>
%	<concept_desc>Applied computing~Law, social and behavioral sciences</concept_desc>
%	<concept_significance>500</concept_significance>
%	</concept>
%	</ccs2012>
%\end{CCSXML}
%\ccsdesc[500]{Applied computing~Law, social and behavioral sciences}

\begin{abstract}
	Electricity theft, the behavior that involves users conducting illegal operations on electrical meters to avoid individual electricity bills, is a common phenomenon in the developing countries. 
	Considering its harmfulness to both power grids and the public, several mechanized methods have been developed to automatically recognize electricity-theft behaviors. 
	However, these methods, which mainly assess users' electricity usage records,  can be insufficient due to the diversity of theft tactics and the irregularity of user behaviors.   

	In this paper, we propose to recognize electricity-theft behavior via multi-source data. 
	In addition to users' electricity usage records, we analyze user behaviors by means of regional factors (non-technical loss) and climatic factors (temperature) in the corresponding transformer area.  
	By conducting analytical experiments, we unearth several interesting patterns: for instance, electricity thieves are likely to consume much more electrical power than normal users, especially under extremely high or low temperatures.
	Motivated by these empirical observations, we further design a novel hierarchical framework for identifying electricity thieves. 
	Experimental results based on a real-world dataset demonstrate that our proposed model can achieve the best performance in electricity-theft detection (e.g., at least +3.0\% in terms of F0.5) compared with several baselines. 
	Last but not least, our work has been applied by the State Grid of China and used to successfully catch electricity thieves in Hangzhou with a precision of \textit{15\%} (an improvement from 0\% attained by several other models the company employed) during monthly on-site investigation. 
\end{abstract}


\maketitle

\section{Introduction}
\label{sec:intro}
\begin{figure}[t]
	\begin{minipage}{0.48\textwidth}
		\centering
		\includegraphics[width=1\textwidth]{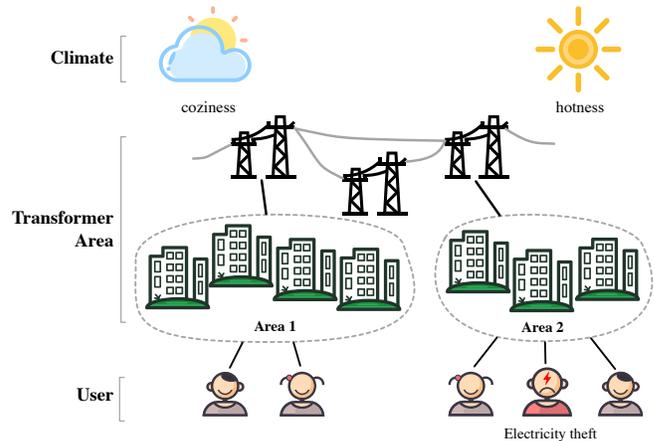}
	\end{minipage}
	\caption{\small An example of electricity theft. The power is supplied to different transformer areas (area 1, area 2). The climate in area 2 is hotter than that in area 1; thus, the power consumption is higher. In order to avoid high electricity bills, several households in area 2 try to pilfer the electricity. \normalsize}
	\label{fig:intro}
\end{figure} 

Electrical power is an important national energy resource \cite{chao1996market}. One of the barriers to the stable provision of electrical power is \textit{electricity theft}.  
Formally speaking, electricity theft refers to the illegal operations by which users unauthorizedly tamper the electricity meter or wires to reduce or avoid consumption costs. 
Electricity theft not only results in unbearable economic losses to the power suppliers, 
but also endangers the safety of electricity users, the electrical systems, and even the public at large. 
As reported in \citep{news-northeast}\footnote{Source report is conducted by Northeast Group LLC.}, 
electricity theft and other so-called ``non-technical loss'' result in a staggering \$96 billion in losses globally per year;
moreover, a shocking statistic is that in 2012,  GDP in India was reported to drop 1.5\%  as a result of electricity theft, while Uttar Pradesh, the most populous state in India, lost 36\% of its total electric power to theft of this kind \cite{india-steal}. 

Great efforts have been made to detect and prevent electricity theft. 
The most intuitive way of doing this is to utilize hardware-driven methods: in short, to find out how the thieves are pilfering the power, then design and upgrade the meter structures accordingly~\citep{depuru2011electricity,guo2010survey, fennell1983pilfer}. 
For instance, \citet{guo2010survey} surveys and summarizes the most commonly used electricity pilfering methods, 
which include changing the structure or wiring mode of the meter; 
then some countermeasures associated with the electrical meters are proposed, including installing a centralized or fully closed meter box for residents. 
However, there are three major drawbacks of these hardware-driven methods: 
1) they require expert domain knowledge to specify the techniques that the electricity thieves use;
2) it is difficult to design a general meter structure, as different regions may often have different pilfering tactics;
3) these methods lose their effectiveness once thieves change their tactics. 

To address these problems, data-driven methodologies have been applied to the task of electricity-theft detection.
Before reviewing the existing works on this topic, it is necessary to explain one associated technical term : \textit{non-technical loss} (NTL)~\cite{chauhan2013non}.
In practice, losses in a utility distribution grid are classified as technical and non-technical.
Technical loss is mainly caused by unwanted effects (e.g., heating of resistive components, radiation, etc.), and is unavoidable\footnote{See \url{https://en.wikipedia.org/wiki/Losses_in_electrical_systems} for details.}.
By contrast, non-technical loss is defined as the energy that is distributed but not billed;
in other words, this type of loss is caused by issues in the meter-to-cash processes.
Although disparate issues may contribute to non-technical losses, a large proportion of the reasons that cause NTL are related to the electrical power pilfering and frauds~\cite{ene-ntl}.
It is therefore straightforward to detect electricity theft from abnormal NTL records~\cite{ahmad2017non,viegas2017solutions}. 
NTL provides transformer-area information. 
To capture individual patterns, many existing work have utilized electricity usage records or NTL as input~\cite{hanmei2004analysis,han2002talking,nagi2009nontechnical,depuru2011support,konstantinos2019efficient,zheng2017wide},
and applied various machine learning techniques (e.g., SVM, CNNs, RNNs) to identify electricity theft. 
However, most of these methods have failed to obtain good performance, due to the diversity and irregularity of electricity usage behavior, which is almost impossible to fully understand either using NTL or electricity usage records alone. 

Motivated by the abovementioned concerns, in this paper, we propose to recognize electricity-theft behavior by bridging three distinct levels of information: 
micro, meso, and macro. 
At the micro- and the meso-level, we seek to capture users' abnormal behavior from electricity usage records and NTL respectively. 
At the macro-level, moreover, we creatively study how climatic conditions influence electricity-theft behavior. 
We then effectively integrate these three levels of information into a uniform framework. 
Our work achieves significant progress in the real-world application of electricity-theft detection 
by deploying the proposed model in State Grid of China\footnote{The state-owned electric utility of China, and the largest utility in the world.} and improving the theft detection performance 

To be more specific, we employ a multi-source dataset, which contains two years' worth of daily electrical power consumption records from 311K users in Zhejiang province, China during June 2017 to April 2019,
along with the corresponding daily NTL records and climatic condition data in all transformer areas covered by those users.  
In addition, we have 4,501 (1.45\%) electricity-theft labels representing 4,626 cases of electricity theft (note that a single user may pilfer electrical power several times within the relevant timespan),
all of which were confirmed during several large-scale on-site investigations conducted by State Grid staff over the two years in question. 
Our empirical observations find that:
1) at the macro-level, climatic condition (or more specifically, \textit{temperature}, which is our main focus on in this paper) affects users' electricity consumption to some extent, 
while users belonging to different groups may show different levels of correlations between electricity usage and temperature;
2) as previous works have pointed out~\cite{ahmad2017non,viegas2017solutions},  NTL is an important factor in detecting electricity thieves at the meso-level;
our detailed analysis shows that abnormal NTL patterns in the transformer areas are a strong signal on electricity theft; 
3) at the micro-level, since electrical power consumption fluctuates with time, we can clearly see the temporal relationships within the individual electricity usage, such that the unusual patterns during a specific period may indicate abnormal user behavior. 
These findings are hierarchically illustrated in \figref{fig:intro}, 
which indicates that users consume more electric power as the temperature gradually increase (due to e.g. the usage of air conditioners).
In order to avoid high electricity bills, some users may employ some tactics to pilfer electricity, which leads to some abnormal fluctuations in their electricity-usage sequences as recorded by the smart meters. 
At the same time, the NTL of the corresponding transformer area can reveal anomalies related to abnormal electricity usage. 

Despite the interesting insights provided by our empirical observation,
the question of how to integrate multi-source information into a uniform framework remains a challenging one.
To capture the temporal and spatial correlations of multi-source sequences, one straightforward method involves first concatenating them at each temporal point, then adopting a single latent representation to capture the overall patterns, such as \textit{multiscale recurrent neural networks} (MPNN) \citep{graves2013speech}.
However, concatenation from different sources widens the feature dimensions, which precludes capturing the significant influences of macro- or meso-level information on micro-level information (\secref{sec:observe}).
Therefore, we propose a hierarchical framework, named \methodname(\methodshort), to extract features and fuse them step by step (\secref{sec:model}).
Experimental results of the electricity-theft detection task on a real-world dataset demonstrate the effectiveness of the proposed \methodshort~method (\secref{sec:exp}). 

Most excitingly, \methodshort~has been employed for real-time electricity-theft detection in the State Grid of China. 
During the monthly on-site investigation in August of 2019, 
we successfully caught electricity thieves in Hangzhou, China, and punished them instantly by checking the suspected samples predicted by \methodshort,
representing an improvement in precision from \textbf{0\%} to \textbf{15\%} (\secref{sec:exp:case}). 
The success of the proposed model in this real-world application further illustrate its validity.
Accordingly, the contributions of this paper can be summarized as follows:
\begin{itemize} [leftmargin=*]
	\item We analyze electricity-theft behaviors from three distinct levels of information based on multi-source data. 
	\item  Based on observational studies, we propose the \methodname (\methodshort) model, which identifies electricity theft by fusing different levels of information, and validate its effectiveness using a real-world dataset. 
	\item We apply \methodshort~ to catch electricity thieves on-site and achieve significant performance improvements. 
\end{itemize}

\section{Problem Definition}
\label{sec:setup}
Let $\mathbf{U}$ be a set of users, while $\mathbf{A}$ is the set of corresponding transformer areas;
that is, each user $u \in \mathbf{U}$ belongs to a specific transformer area $a \in \mathbf{A}$ based on the regional location. An area $a$ usually contains hundreds of users.
Each user $u$ has electricity usage records with $T$ observations within a certain timespan, which we refer to as $\mathcal{X}^e_u = \{\mathbf{x}^e_1, ..., \mathbf{x}^e_{T}\}$. 
Each area $a \in \mathbf{A}$ has the NTL records, denoted as $\mathcal{X}^l_a= \{\mathbf{x}^l_1, ..., \mathbf{x}^l_{T}\}$, 
and the observation sequence of climatic conditions, written as $\mathcal{X}^c_a = \{\mathbf{x}^c_1, ..., \mathbf{x}^c_{T}\}$. 
The denotation $\mathbf{x}^*_t \in \mathbb{R}^{d_*}$ represents different quotas with various dimensions (see details in \secref{sec:observe:data}). 
In light of the above, we can define the problem addressed in this paper as follows:  

\begin{definition}
	\textbf{\textit{Electricity-theft detection.}} 
	Given a specific user $u$ who belongs to the transformer area $a$, the goal is to estimate $\mathbf{P}(\mathcal{Y} | \mathcal{X}^e_u, \mathcal{X}^l_a, \mathcal{X}^c_a)$, which is the probability that $u$ pilfers electrical power ($\mathcal{Y}=1$) or not ($\mathcal{Y}=0$). 
\end{definition} 

In the following sections, 
we will conduct observational studies on the multi-source sequences to obtain potential insights.
Based on these intuitions, we then propose a novel hierarchical framework for recognizing electricity-theft behavior.

\section{Empirical observations}
\label{sec:observe}
Although existing data-driven electricity-pilfering-detection methods seek to capture the characteristics of users' electrical power consumption,
it is rather difficult in real-world applications to efficiently catch electricity thieves if we only observe the user's electrical power consumption records;
this is due to the diversity, complexity and irregularity of electricity-theft behavior.
Therefore, we compile a multi-source dataset, with additional NTL and temperature records for each transformer area,  in order to analyze the traces of users' electricity usage.
The associated observational findings and insights are presented in this section.

\subsection{Dataset Description}
\label{sec:observe:data}
Our dataset comprises three parts:
the two sets of electricity-related records are provided by State Grid Zhejiang Power Supply Co. Ltd.\footnote{http://www.sgcc.com.cn/ywlm/index.shtml},
while the temperature records are collected from the official weather website.
The overall statistics are summarized in \tableref{tb:observe:statistics}.

\vpara{Electricity Usage Records.} 
This dataset covers the daily electrical power consumption records of \textbf{310,786} users in total, 
ranging from June 2017 to April 2019.
For each user, we have the total, on-peak and off-peak electricity usage (kW$\cdot$h) records for each day within the relevant timespan.

\vpara{Non-Technical Loss Records.}
This dataset contains the daily meso-level electrical records from a total of \textbf{3,908} transformer areas, covering all of these 311K users, 
and has the same time range as the usage dataset. 
More specifically, for each area,  
the daily amount of electrical power (kW$\cdot$h) lost due to  non-technical loss (NTL) is recorded.

\vpara{Temperature Records.} 
We obtained the temperature records for all prefecture-level cities in Zhejiang
during the same timespan as above from the Weather Radar\footnote{http://en.weather.com.cn}.
For each city, these records contain the maximum and minimum temperatures (\textcelsius{}) for each day.

\begin{table}[b]
	\centering
	\caption{Overall statistics of the datasets.}
	\label{tb:observe:statistics}
	\begin{tabular}{>{\centering}p{4.8cm}c}
		\toprule
		Metric & Statistics \\
		\midrule
		$\#$(users) & 310,786\\
		$\#$(transformer areas) & 3,908\\
		$\#$(electricity thieves) & 4,501\\
		$\#$(pilfering-cases) & 4,626\\
		$\#$(prefecture-level cities) & 11\\
		\bottomrule
	\end{tabular}
\end{table}

\vpara{Labels of Electricity Thieves.}
Among all users, \textbf{4,501 (1.45\%)} were confirmed by State Grid staff to be \textit{electricity thieves} during their on-site investigations;
it should be noted that there are a total of \textbf{4,626} electricity-pilfering cases, 
since a single user may be caught committing theft several times during the two years.
We then regard all remaining users (98.55\%) as normal, 
i.e., as having engaged in no pilfering behavior over the entire timespan. 
While it is possible that a few users who adopt subtle ways to pilfer electricity, which are not caught; these cases bring in the noise but are very rare. 
We take these confirmed cases as ground-truth and collect the timestamps when electricity thieves were caught for detailed analysis and experiments.

\subsection{Micro-level: User Observation}
\label{sec:observe:user-level}
\begin{figure}[t]
	\begin{minipage}{0.48\textwidth}
		\centering
		\includegraphics[width=1.0\textwidth]{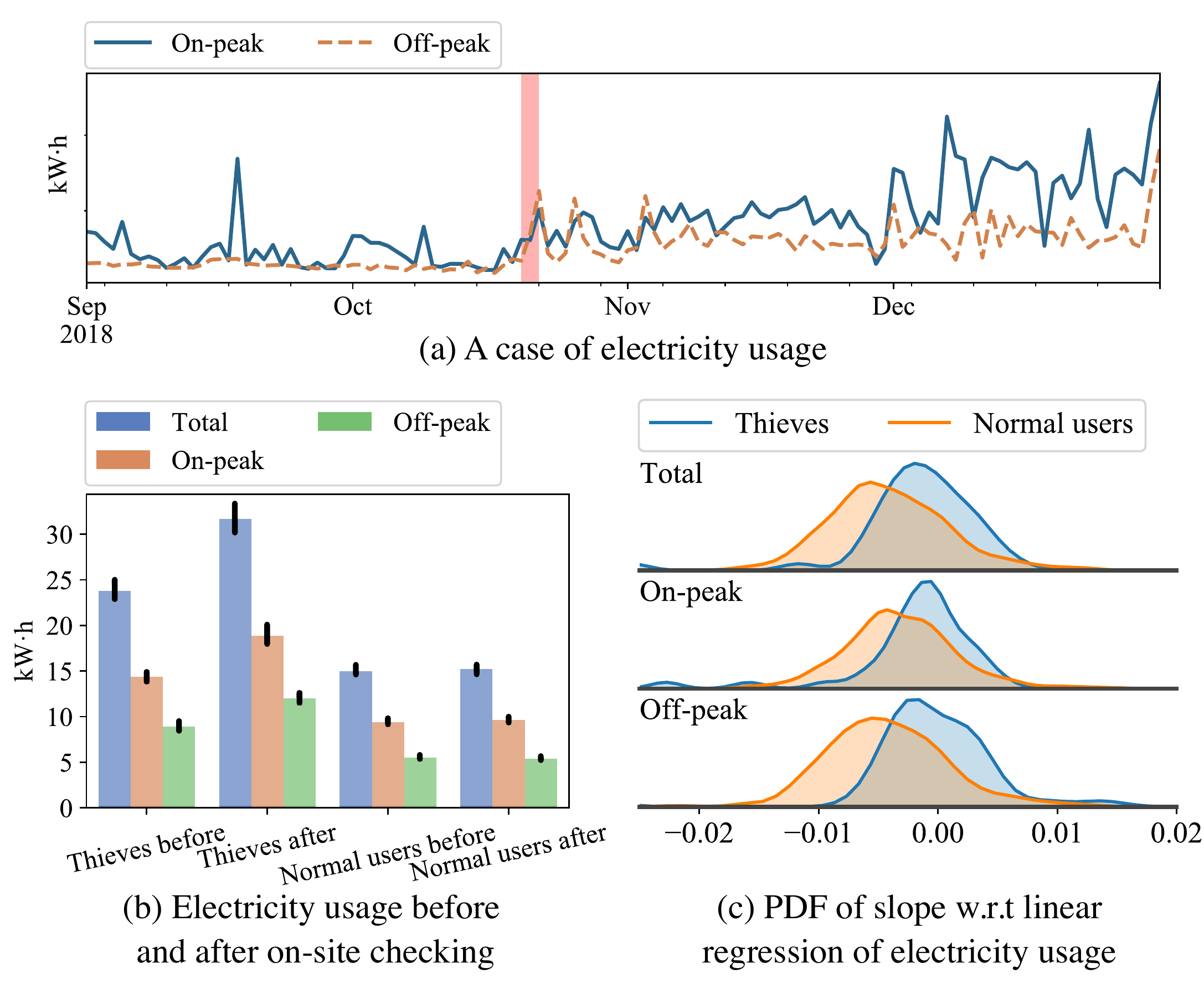}
	\end{minipage}
	\caption{\small A study of micro-level factors and electricity theft. (a) presents a case of electricity usage with the red bar indicating the time at which the thief was caught. (b) presents the statistics of all users pertaining to electricity usage before and after on-site checking, while the trends of the usage are shown in (c). We can therefore see that the behaviors of electricity thieves are different from those of normal users. \normalsize}
	\label{fig:observation:user}
\end{figure} 
We first examine the micro-level electricity usage records to see 
whether abnormal patterns or characteristics in the user behavior exist that indicate electricity theft.

To better understand pilfering behaviors, we first have to recognize the time periods during which the thieves were pilfering electricity.
According to the timestamps of on-site investigations, 
the records for each electricity theft can be divided into two parts: before and after being caught.  
For convenience we refer to the specific timestamp as a \textit{checkpoint}. 
We further assume that a thief was stealing the power during the last 30 days before the checkpoint,
then returning back to a normal state in the following 30 days after the checkpoint.
This is based on the domain knowledge that we have chosen a month as the timespan, 
since an electricity thief often steals power for a long time (probably for a period longer than one month), 
and once caught and punished, he/she would instantly cease engaging in pilfering behavior and behave normally.
To give a concrete example, \figref{fig:observation:user}a presents a case of an electricity thief 
who was caught in the middle of October 2018 (red bar). 
We can see an abnormal pattern in the middle of September, 
showing that he had a sudden peak of electricity usage,
while he had almost no off-peak electrical power usage during Sep. and Oct. 
Moreover, in this case, there is no significant decreasing trend of electricity usage before he was caught;
in practice, however, users would probably use less electricity in Sep. and Oct. due to the temperature drop.
These clues may revel the thief's abnormal state or behavior. 

In light of the above assumption and case study, we provide an overview of the electricity consumption of both electricity thieves and normal users in \figref{fig:observation:user}b, \ref{fig:observation:user}c.
We draw the distributions of daily power usage at different times (before and after being caught) in \figref{fig:observation:user}b;
for each group of settings, we show the average value along with the standard error bar.
We can clearly see that for electricity thieves, 
there is a significant increment of electricity usage after being caught compared with before (the two histograms on the left-hand side).
This is reasonable, since pilfering behavior would reduce the electricity that they actually consume.
An opposite trend is exhibited by the normal users in that the usage before and after the checkpoint seems unchanged;
in other words, the characteristics of the electrical power consumption behaviors of normal users are relatively stable.

Another obvious observation from \figref{fig:observation:user}b is that 
electricity thieves exhibit a much higher level of electricity usage compared with normal users,
although pilfering behavior would cut down the recorded value of electrical power consumption. 
One reasonable explanation for this phenomenon comes from the idea that people typically engage in risky behavior only when they expect a higher benefit.
As for this scenario, users whose electricity consumption is high are expected to have a far stronger motivation to steal power,
as they would reduce their costs significantly through engaging in pilfering behavior.
By contrast, if a user uses a very small amount of electricity, 
there is no need for him/her to undertake such illegal operations
since this behavior would also be economically foolish, 
as once being caught, the user would be fined vast amounts.

To further illustrate the differences in the stability of electricity usage between electricity thieves and normal users, 
we conduct a linear regression on electricity usage during the three months before the theft was identified. 
For better visualization and explanation, we set a restriction on the checkpoint: 
in short we only sample the thieves who were caught during October.
Under these settings, user are expected to reduce electricity usage stably at the end of Aug. and Sep. compared with Jul. due to the drop in temperature, 
such that the linear regression may be able to capture this decreasing trend.
We use a probability density function (PDF) \citep{parzen1962on}  to compare the distributions of slopes in the linear regression. As \figref{fig:observation:user}c shows, the gap of PDF curves between electricity thieves and normal users represent the different trends in electricity usage.
More specifically, and in line with our commonsense expectations, 
the usage of the majority of normal users (80.30\%) exhibits a negative slope;
by contrast, for electricity thieves, the peak of the PDF lies around 0, 
while nearly half of them even have a positive slope, which is entirely opposite to the common cases.

Combined with the analysis mentioned above, we can conclude that 
electricity thieves have a higher level of electrical power consumption with less stability.

\subsection{Meso-level: Area Observation}
Although we can observe several differences between electricity thieves and normal users in terms of electricity usage, 
due to the fact that user behaviors are very complicated and may be affected by many factors,
focusing only on those micro-level differences prevents us from efficiently identifying electricity thieves. 
Here, another intuitive element of knowledge is that the transformer area records the overall electricity consumption of different users, i.e., non-technical loss (NTL). Hence, in order to reveal the meso-level characteristics of electricity thieves, 
we conduct an analysis combining regional information with individual electricity usage.

\begin{figure}[t]
	\centering
	\begin{minipage}{0.48\textwidth}
		\includegraphics[width=1.0\textwidth]{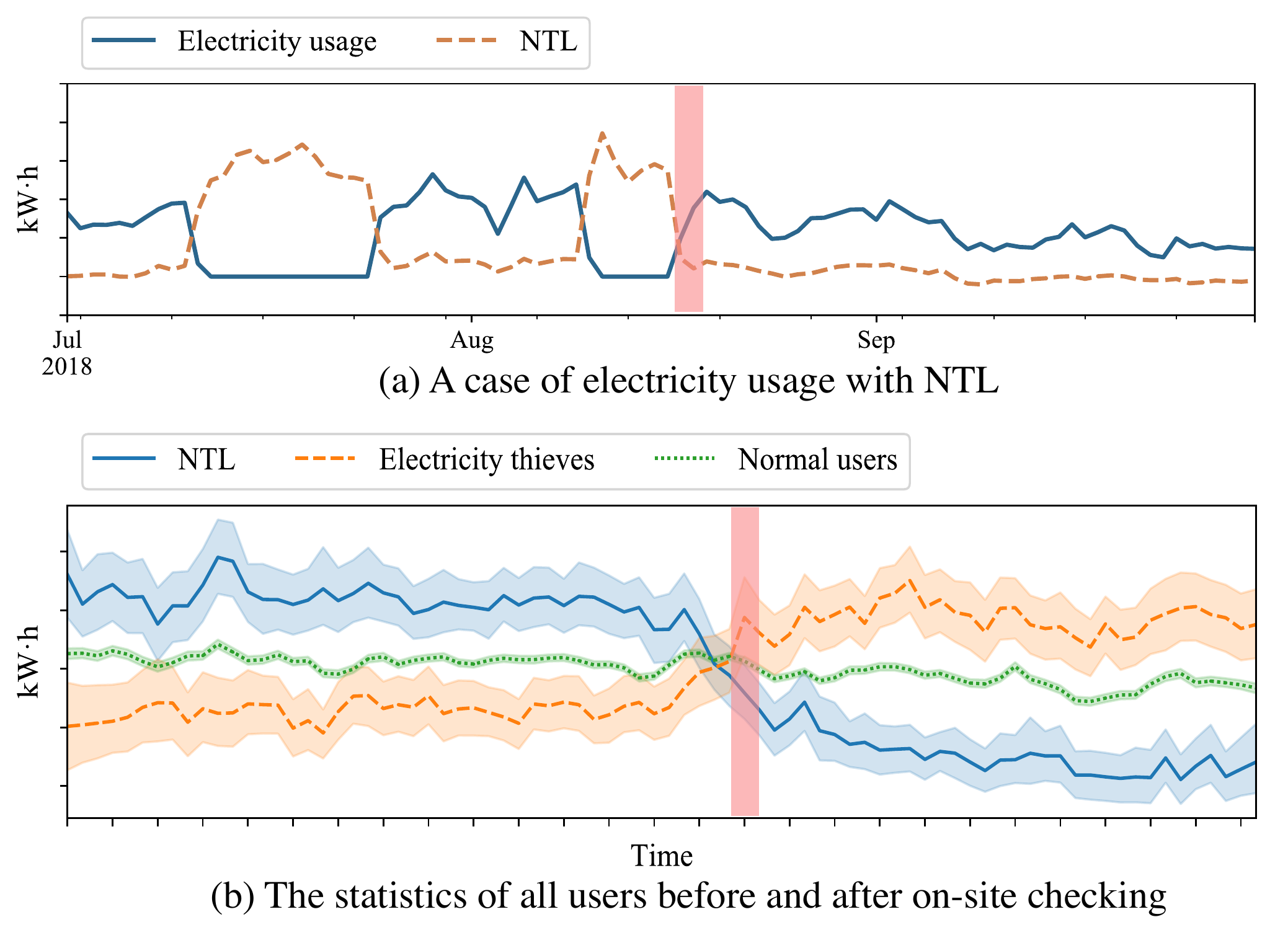}
	\end{minipage}
	\caption{\small The study of meso-level factors to electricity theft. (a) presents an electricity-usage case with NTL and the red bar indicates the time of being caught as thief. (b) presents the statistics of all users and transformer areas. Both of them illustrate that the abnormal increment of NTL can be caused by pilfering electricity. \normalsize}
	\label{fig:observation:loss}
\end{figure} 

We begin with a real-world case study of an electricity theft in \figref{fig:observation:loss}a.
The user was caught as a thief in the middle of August (red bar).
A clear and interesting observation is that the trend of correlations between the user's electricity usage and NTL before he was caught
is completely opposite to that of the period after on-site checking was conducted. 
More specifically, before the user was caught, the less electricity usage his records showed (probably a large part of the power used was stolen),
the higher the NTL of the transformer area would be;
after on-site checking, however, the NTL of the area became relatively stable.

We can confirm this finding on the whole dataset by observing the correlations between the averaged NTL of all transformer areas and the individual electricity usage of all users during the period before and after on-site checking (\figref{fig:observation:loss}b).
In more detail, we move the sliding window to retrieve the electricity usage of each thief:
1) 50 days before, and 2) 30 days after the checkpoint.
We then sample the usage records of all normal users in the same transformer area and during the same period as the thief.
Again, we can see that the electricity usage of normal users (green line) are rather stable during the observed timespan;
for electricity thieves (orange line), however, 
their averaged power consumption significantly increased once they were caught, 
while the NTL of their transformer areas dropped accordingly. 
It gives us the clue that the NTL in the transformer area may be a signal for 
indicating whether there are electricity thieves in this area,
and additionally, 
that capturing such correlations could be helpful in electricity-theft detection.

\subsection{Macro-level: Climate Observation}
\begin{figure}[t]
	\begin{minipage}{0.48\textwidth}
		\centering
		\includegraphics[width=0.8\textwidth]{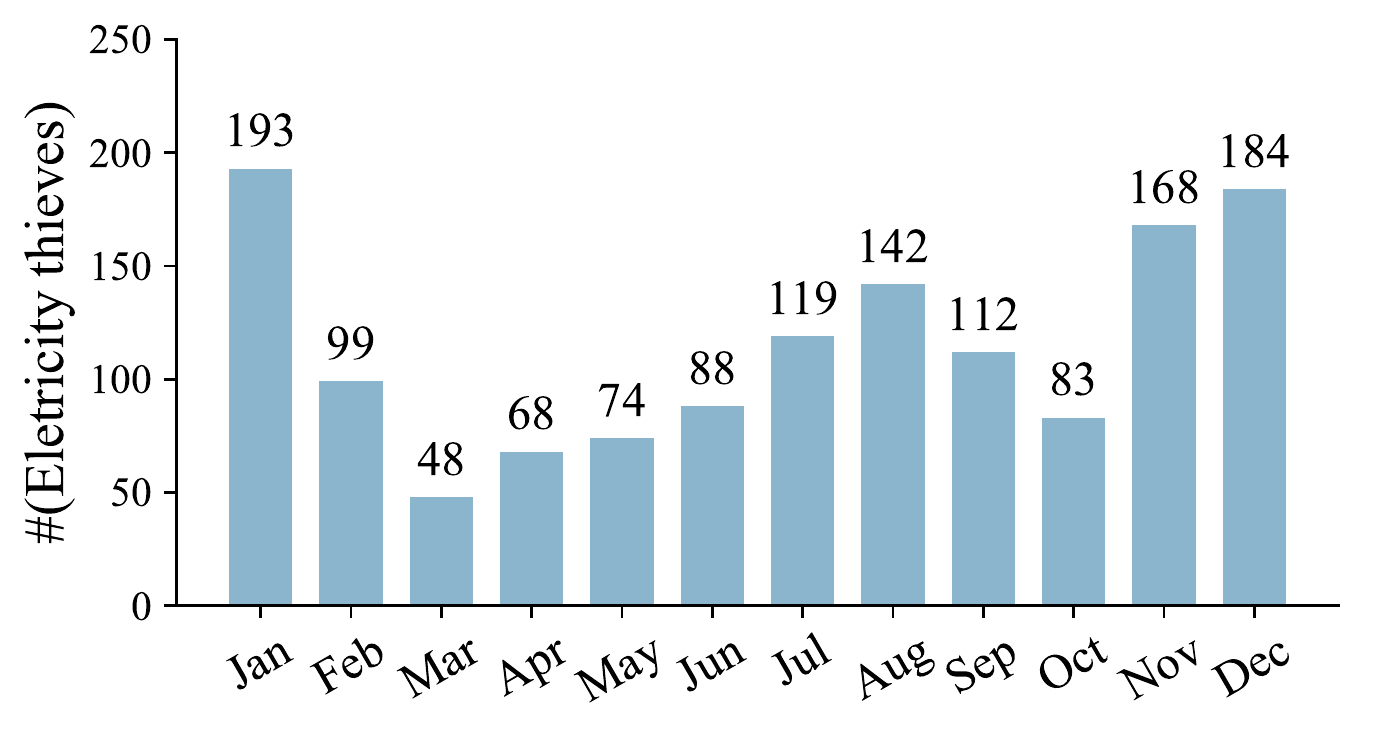}
	\end{minipage}
	\caption{\small Number of electricity thieves in each month.\normalsize}
	\label{fig:steal_num}
\end{figure}

Previous work~\cite{kamga2013hailing} has demonstrated the influence of climatic conditions on user behavior (taxi ordering). 
It would be interesting to determine whether some relationship between climate and electricity usage exists in our scenario, 
or whether the macro-level factors affect users' electrical power usage in a non-trivial way. 
Accordingly, we present the statistics of the seasonal effects on electricity-pilfering behavior (\figref{fig:steal_num}), 
which reveal that  most pilfering cases were caught in winter (Dec., Jan.) and summer (Jul., Aug.).
This leads to the straightforward conclusion that the climatic conditions would influence the user behavior of electricity usage, especially for the electricity thieves. Inspired by the practical experiences, temperature is the only climatic variable we consider in the present research,
since the most obvious difference between winter and summer is the temperature factor, 
and it is believed that temperature will influence users' daily behavior in an intuitive way.
We use the averaged value of maximum and minimum in a day to represent the daily temperature;
this setting is used throughout the entire paper unless otherwise indicated.

\begin{figure}[t]
	\centering
	\begin{minipage}{.48\textwidth}
		\includegraphics[width=1.0\textwidth]{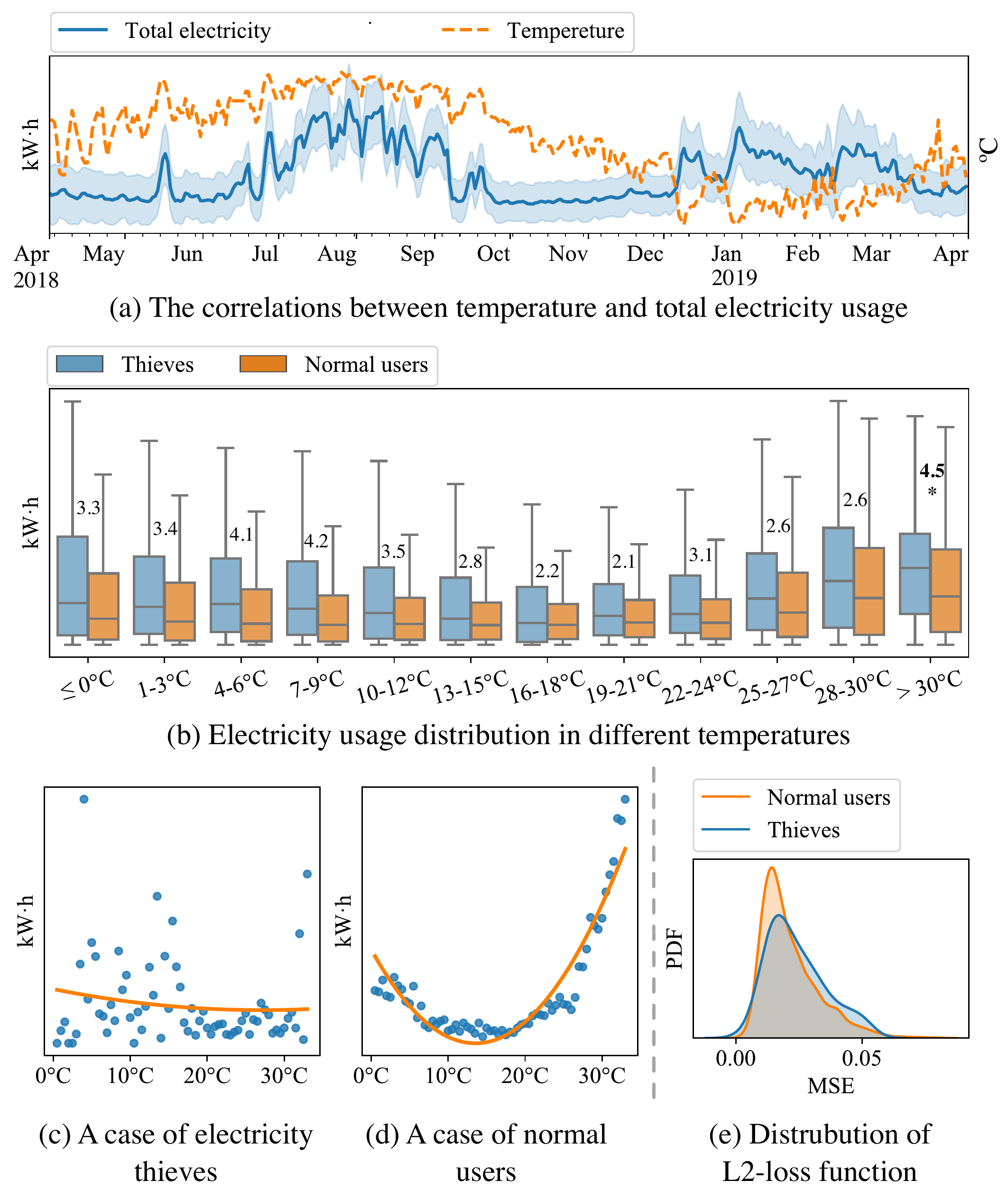}
	\end{minipage}
	\caption{\small The study of macro-level factors to electricity theft. (a) presents the strong correlations from climatic conditions to user behaviors. (b)-(e) illustrate the electricity-usage irregularity of thieves under different temperature conditions. The floating numbers in (b) indicate the Wasserstein distance of distribution between the thieves and normal users. \normalsize}
	\label{fig:temperature}
\end{figure}

We first illustrate that a correlation does indeed exist between temperature and users' electricity usage,
as shown in \figref{fig:temperature}a.
We indicate the averaged daily total electricity consumption of all users over one year with an error bar (blue line), 
while the orange line indicates the temperature each day.
Total electricity usage fluctuates with the temperature, as some sharp peaks and valleys are coincident; 
the most obvious correlations between these two factors is that 
extremely high or low temperatures are associated with increased electricity consumption.
This would appear to be a commonsense observation since extreme temperatures are of course associated with the increased usage of high-powered appliances such as air conditioning or heaters;
here we verify this assumption by means of a brief visualization. 

We next examine the relationships between temperature and electricity usage among different groups of users.
In \figref{fig:temperature}b, we aggregate the daily total electricity usage based on the temperature of that day, 
then draw a boxplot with respect to our two user groups of interest, i.e. electricity thieves and normal users.
In line with what we have observed in \secref{sec:observe:user-level},
we again find that electricity thieves have a much higher level of electrical power consumption regardless of the temperature.
Furthermore, additional observation reveals the specific influence of temperature on electricity pilfering,
as the gap of daily total electricity usage between thieves and normal users is significantly larger under extremely high or low temperatures compared with average temperatures:
if we regard a temperature lower than 0\textcelsius{} or higher than 30\textcelsius{} as extreme conditions (as is accepted by the public), 
the average gap of daily total electricity usage between thieves and normal users under extreme conditions is  4.51kW$\cdot$h, 
while that for non-extreme conditions (between 0 and 30\textcelsius{}) is 3.03kW$\cdot$h (a decrease of 32.8\%); 
moreover, if we restrict the temperature span to between 10\textcelsius{} and 25\textcelsius{}, the average gap becomes 2.70kW$\cdot$h.
We further compute the Wasserstein distance ($d_w$)~\citep{cuturi2014fast} of electricity usage distributions between thieves and normal users under different temperatures,
and find that distances for extreme temperatures ($d_w$=4.5 ($>$30\textcelsius{})) are much larger than other cases (e.g., $d_w$=2.1 ($19-21$\textcelsius{})).
This verifies our conclusion from a statistical perspective that
electricity thieves are likely to consume much more electrical power than normal users, 
especially under extremely high or low temperature conditions.

Finally, we take a deeper look into these correlations by means of several case studies.
We choose a typical electricity thief along with a normal user in the same transformer area as examples,
then show the pairs of daily total electricity and temperature in the form of 2-D figures.
For better visualization and illustration, we also draw the second-order regression curve in both \figref{fig:temperature}c and \ref{fig:temperature}d.
Again, the trends of the normal user's curve is consistent with our experience suggesting that 
very high or low temperatures are associated with elevated electricity consumption,
and the points are well fitted to a specific quadratic curve (\figref{fig:temperature}d);
by contrast, the scatterplot for the case of the electricity thief seems disordered (\figref{fig:temperature}c). 
It is better to explore such correlations by merging users together within different groups;
however,  since the scale of each user's electricity consumption may vary substantially,
the scatterplot of the data for all normal users may not well fit to a quadratic curve.
An alternative approach would to individually fit the points for each user, 
then show the gap in the distributions of the fitting loss between normal users and electricity thieves (\figref{fig:temperature}e).
The two PDFs simply reveal the fact that scatters of normal users can be fitted more easily to a quadratic curve than thieves. 

Summarizing from the abovementioned observations,
we can conclude that the analysis based on the combination of electricity usage, NTL and temperature can yield extra information related to electricity-theft behavior, 
and can also give us strong insights and motivations to consider the correlations among these three different levels of factors when detecting electricity theft.

\section{Our approach}
\label{sec:model}
In this section, we integrate the insights gained from empirical observations (\secref{sec:observe}) into a hierarchical framework, named \methodname (\methodshort),  to capture the behavioral patterns from multi-source observation sequences.

\vpara{Overview.}
Motivated by \secref{sec:observe}, we model the user behaviors based on three distinct levels of information,  as follows:
\begin{itemize} [leftmargin=*]
	\item \textit{Macro-level}. We define $\mathcal{X}^c_a, a\in\mathbf{A}$, to represent the observation sequence of temperature, 
	which reflects the climatic conditions that may influence users' electricity consumption patterns.
	\item \textit{Meso-level}. We define $\mathcal{X}^l_a, a\in\mathbf{A}$, to represent the observation sequence of non-technical loss (NTL), which indicates the real-time status of the corresponding transformer area.
	\item \textit{Micro-level}. We define $\mathcal{X}^e_u, u\in\mathbf{U}$, to represent the observation sequence of users' electricity usage, which presents the trace of individual electrical power consumption behaviors.
\end{itemize}
The empirical observations demonstrate that the macro- and meso-level information influence the micro-level behaviors to some extent.
In order to integrate the multi-source information so as to capture the abnormal behavioral patterns of electricity thieves, intuitively, we develop a hierarchical framework to extract features from their respective different sources,  and fuse them step by step.

\begin{figure}[tb]
	\begin{minipage}{0.48\textwidth}
		\raggedleft
		\includegraphics[width=0.96\textwidth]{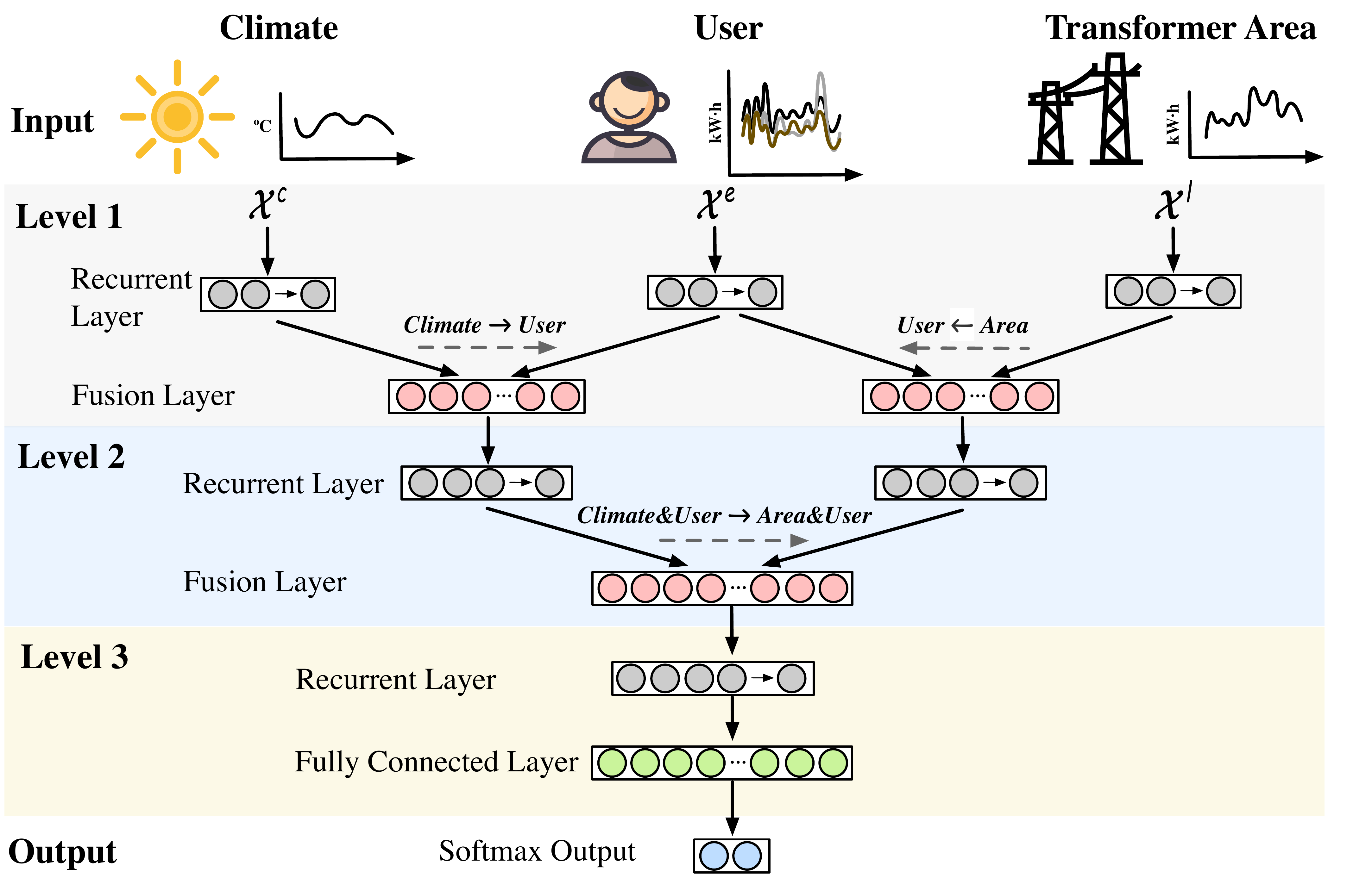}
	\end{minipage}
	\caption{\small Architecture of \methodshort.  Given a multi-source observation sequence (climate, area and user), \methodshort~constructs a three-level framework: at each level, different sequences are inputted into different recurrent layers respectively, the latent representations of which are fused in pairs. Dashed lines indicate the direction of the fusion. The last layer outputs the probability of electricity theft.
		\normalsize}
	\label{fig:model}
\end{figure}  

\subsection{Model Description}
\label{sec:model:define}
When modeling the multi-source sequences, a straightforward baseline can be established by first concatenating them at each temporal point, then using a single latent representation to capture the overall patterns, such as \textit{MRNN} \citep{graves2013speech}. 
However, concatenation from different sources widens the feature dimensions, which may preclude capturing the significant correlations between different levels of information.
More specific, in our scenario, different users in the same transformer area may have the same observation sequence of NTL or temperature; therefore, the straightforward concatenation could result in a confusion of user behaviors from the regional and climatic levels. 
Accordingly, we opt to better extract the independent features of each source respectively, and then conduct the pairwise information fusion.

In \figref{fig:model}, we present the architecture of the proposed \methodshort framework. 
In addition to input and output, \methodshort contains three levels of feature extraction and hierarchical fusion, 
which aim to capture different levels of influence between data sources. 
Each level is described in more detail below.
\begin{itemize} [leftmargin=*]
	\item \textbf{Level 1:} Captures the temporal patterns in the observation sequence independently (e.g. the patterns in temperature ($\mathbf{h}^c$), NTL ($\mathbf{h}^l$) and user's electricity usage ($\mathbf{h}^e$)), and fuses them in pairs ($\mathbf{h}^c \rightarrow \mathbf{h}^e$, $\mathbf{h}^l \rightarrow \mathbf{h}^e$). It aims to model the influence from macro- or meso-level factors on user behaviors respectively.
	\item \textbf{Level 2:} Captures the temporal patterns after preliminary fusion at Level 1 (e.g. user-climate ($\mathbf{h}^{ec}$) and user-area ($\mathbf{h}^{el}$)) respectively, and then fuses the patterns of $\mathbf{h}^{ec} \rightarrow \mathbf{h}^{el}$. It aims to uniform the influence from macro and meso level on user behaviors.
	\item \textbf{Level 3:} Captures the overall temporal patterns in the multi-source sequence ($\mathbf{h}^{ecl}$). The hierarchically fused information is integrated to capture the behavioral patterns, which can be applied to estimate the probability of electricity theft.
\end{itemize} 

\noindent Note that we do not fuse the representations $\mathbf{h}^c$ and $\mathbf{h}^l$ in the first level. This is because our observational studies (\secref{sec:observe}) suggest that these two distinct levels of information are uncorrelated, although they are closely related to user's electricity usage.
Here, we aim to capture how these two factors influence user electricity consumption behavior.

Based on the hierarchical construction, \methodshort~can conduct feature extraction and fusion in each level and gradually integrate the information between multiple sources.
As for the operations in each level, we define a uniform formulation: given the sequential input $\mathbf{I}^{(k)}_t$  and another source $\mathbf{I}'^{(k)}_t$ in the $k$-th level, the feature extraction and hierarchical fusion can be formulated as follows:
\begin{equation}
\begin{split}
\mathbf{h}^{(k)}_t &= \mathcal{F}_{\rm{recurrent}}\left(W_{\mathbf{I}\rightarrow \mathbf{h}} \cdot \mathbf{I}^{(k)}_t, \mathbf{h}^{(k)}_{t-1}\right), \\
\mathbf{h}'^{(k)}_t &= \mathcal{F}_{\rm{recurrent}}\left(W_{\mathbf{I}'\rightarrow \mathbf{h}'} \cdot \mathbf{I}'^{(k)}_t, \mathbf{h}'^{(k)}_{t-1}\right), \\
\mathbf{I}^{(k+1)}_t &= \bm{\alpha}_t \cdot \mathcal{F}_{\rm{act}} \left(\mathcal{F}_{\rm{fuse}}\left(\mathbf{h}^{(k)}_t, W_{\mathbf{h}'\rightarrow \mathbf{h}} \cdot \mathbf{h}'^{(k)}_t\right)\right)
\end{split}
\label{eq:model:define}
\end{equation}
$\mathbf{h}^{(k)}_t$ denotes the latent representation of $\mathbf{I}^{(k)}_t$ in the $k$-th level at temporal point $t$, 
which is updated by function $\mathcal{F}_{\rm{recurrent}}$ based on its previous memory $\mathbf{h}^{(k)}_{t-1}$ and current input $\mathbf{I}^{(k)}_t$,
while the apostrophe ($'$) denotes the same meaning of the sequences from another source.
$W_*$ are the trainable weighted matrices, and the function $\mathcal{F}_{\rm{fuse}}$ aims to fuse the information from $\mathbf{I}'^{(k)}_t$  into the current sequence $\mathbf{I}^{(k)}_t$, 
then outputs the intermediate representations by means of an activation function ($\mathcal{F}_{\rm{act}}$).
Moreover, $\bm{\alpha}_t$ denotes the attention coefficient from $\mathbf{h}'_{t}$ to $\mathbf{h}_t$, 
which tries to automatically discover the attention weights from $\mathbf{I}'$ to $\mathbf{I}$ based on ``end-to-end'' learning. 
We will introduce the details of model inference and learning for detecting electricity thieves in the next section.

\begin{figure}[tb]
	\begin{minipage}{0.48\textwidth}
		\centering
		\includegraphics[width=0.78\textwidth]{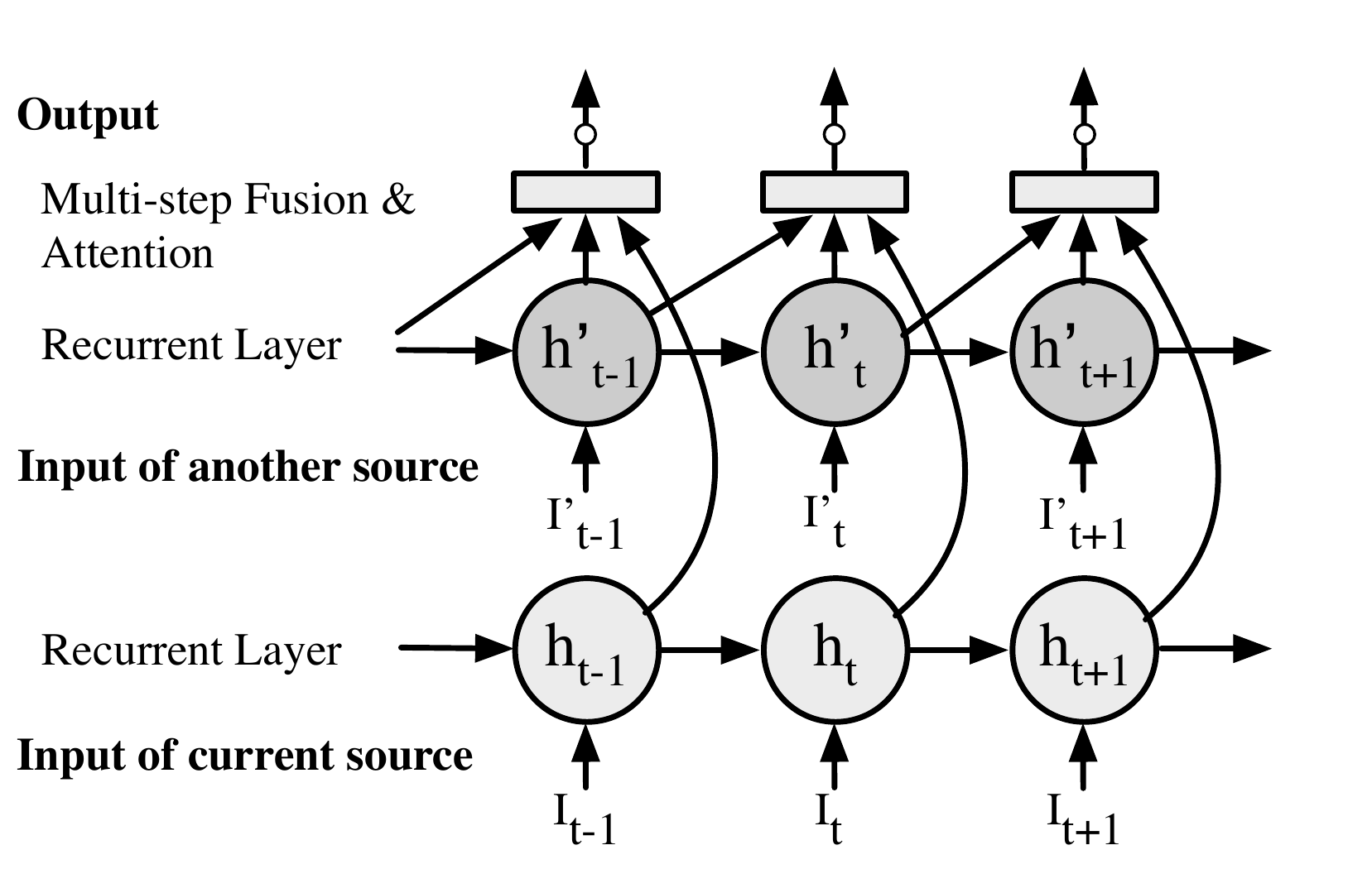}
	\end{minipage}
	\caption{\small The architecture of recurrent and fusion layers on multi-source sequences.  The sequential inputs of another source $\mathbf{I}'$ are integrated into the current ones $\mathbf{I}$ by means of the above architecture, which is a uniform component in \figref{fig:model}. \normalsize}
	\label{fig:review}
\end{figure} 

\subsection{Model Inference and Learning}
Herein, we adopt neural networks to implement \methodshort, the parameters of which are learned by minimizing some specific loss. As for capturing the temporal patterns, we can apply several existing methods to implement the recurrent layer $\mathcal{F}_{\rm{recurrent}}$ (see details in \tableref{tb:exp:rnn} of \secref{sec:exp:effect}). 
As for the fusion function $\mathcal{F}_{\rm{fuse}}$, we propose a new hierarchical fusion mechanism, containing multi-step fusion and attention operations, to effectively bridge the information from different sources.
More details are introduced in the next paragraph.

\vpara{Hierarchical fusion mechanism.}
In order to capture the correlations between different levels of information, we propose a multi-step fusion mechanism. 
The intuition here is that the influence between two distinct levels of sequences may be time-delayed.
For instance, in addition to being affected by today's temperature, a user's electricity usage is also probably related to yesterday's weather;
one concrete example is that if the previous day was very hot, people will tend to turn on the air conditioning for a longer time, even if today is colder. 
Therefore, we should try to fuse more information in the temporal interval rather than just at the current temporal point.
More specifically, as shown in \figref{fig:review}, the current latent representation $\mathbf{h}_t \in \mathbb{R}^{d_\mathbf{h}}$ and that from another data source $\mathbf{h}'_t \in \mathbb{R}^{d_{\mathbf{h}'}}$ are fused via the following formulation:
\begin{equation}
\mathcal{F}_{\rm{fuse}}\left(\mathbf{h}_t, \mathbf{h}'_t\right) = \left(\mathbf{h}_t \odot W_{\mathbf{h}'\rightarrow \mathbf{h}} \mathbf{h}'_{t-1} \right) \oplus \left(\mathbf{h}_t \odot W_{\mathbf{h}'\rightarrow \mathbf{h}} \mathbf{h}'_t \right)
\end{equation}
where $\oplus$ denotes the concatenation operator and $\odot$ is the pooling operator. 
For the $t$-th time step, $\left(\mathbf{h}_t \odot W_{\mathbf{h}'\rightarrow \mathbf{h}} \mathbf{h}'_{t-1} \right)$ and $\left(\mathbf{h}_t \odot W_{\mathbf{h}'\rightarrow \mathbf{h}} \mathbf{h}'_t \right)$ capture how $\mathbf{h}'_{t-1}$ and $\mathbf{h}'_t$ respectively influence $\mathbf{h}_t$. 

However, it is impossible that the fused information at each time step will be equally important to behavioral patterns. 
For example, someone is probably an electricity theft if he consumes little electrical power in summer or winter,
but this may not be true in autumn or spring, since users typically consume less electricity during these months due to seasonal effects.
\citet{liu2019imputation} suggests that models should try to measure such significant information at different temporal points. 
Hence, we design an attention mechanism to model the varying significance of the fused information at different time steps:
given the current fusion level $k$, the input of the next level $\mathbf{I}^{(k+1)} \in \mathbb{R}^{T \times 2d_{\mathbf{h}^{(k)}}}$ is computed by a linear combination of the intermediate representations, weighted by a score vector $\bm{\alpha}^{(k+1)} \in \mathbb{R}^T$:
\begin{equation}
\begin{split}
\mathbf{I}^{(k+1)} = \sum_{t=0}^{T} \bm{\alpha}_t^{(k+1)} \cdot \rm{tanh}\left(\mathcal{F}_{\rm{fuse}}\left(\mathbf{h}^{(k)}_t, \mathbf{h}'^{(k)}_t\right)\right) &\\ 
\bm{\alpha}^{(k+1)} = \rm{softmax}\left(W_{\mathbf{h}\rightarrow \bm{\alpha}} \cdot \sum_{t=0}^{T}  \mathcal{F}_{\rm{fuse}}\left(\mathbf{h}^{(k)}_t, \mathbf{h}'^{(k)}_t\right)\right) &
\end{split}
\end{equation}

\noindent where $\sum$ denotes the concatenation, 
and $W_{\mathbf{h}\rightarrow \bm{\alpha}}$ is a trainable weighted matrix shared by all temporal points. 
The activation function $\rm{tanh}$ is used to activate the intermediate representations.

\vpara{Model formulation.}
So far, we can materialize \equationref{eq:model:define} for \methodshort~by the abovementioned definitions,  
the complete mathematical formulations for which are as follows:
\begin{subequations}
	\small
	\begin{align}
	\mathbf{I}^e &= \mathcal{X}^e_u, \quad \mathbf{I}^l = \mathcal{X}^l_a, \quad \mathbf{I}^c = \mathcal{X}^c_a, \qquad u\in \mathbf{U}, a\in \mathbf{A} \\
	\label{eq:model:all:1}
	\left[ \begin{aligned} \mathbf{h}^e_t \\ \mathbf{h}^l_t \\ \mathbf{h}^c_t \end{aligned} \right] &= \left[
	\begin{aligned}
	\mathcal{F}_{\rm{recurrent}}\left(\mathbf{I}^e_t, \mathbf{h}^e_{t-1}\right) \\
	\mathcal{F}_{\rm{recurrent}}\left(\mathbf{I}^l_t, \mathbf{h}^l_{t-1}\right) \\
	\mathcal{F}_{\rm{recurrent}}\left(\mathbf{I}^c_t, \mathbf{h}^c_{t-1}\right)
	\end{aligned}
	\right], 
	\left[ \begin{aligned} \mathbf{I}^{el}_t \\ \mathbf{I}^{ec}_t \end{aligned}\right] = \left[ \begin{aligned} \bm{\alpha}^{el}_t \\ \bm{\alpha}^{ec}_t \end{aligned}\right] \cdot \rm{tanh}\left[ \begin{aligned}
	\mathcal{F}_{\rm{fuse}}\left(\mathbf{h}^e_t,  \mathbf{h}^l_t\right) \\
	\mathcal{F}_{\rm{fuse}}\left(\mathbf{h}^e_t,  \mathbf{h}^c_t\right)
	\end{aligned}\right] \\
	\label{eq:model:all:2}
	\left[ \begin{aligned} \mathbf{h}^{ec}_t \\ \mathbf{h}^{el}_t \end{aligned} \right] &= \left[
	\begin{aligned}
	\mathcal{F}_{\rm{recurrent}}\left(\mathbf{I}^{ec}_t, \mathbf{h}^{ec}_{t-1}\right) \\
	\mathcal{F}_{\rm{recurrent}}\left(\mathbf{I}^{el}_t, \mathbf{h}^{el}_{t-1}\right)
	\end{aligned}
	\right], 
	\mathbf{I}^{elc}_t = \bm{\alpha}^{elc}_t \cdot \rm{tanh} \left( \mathcal{F}_{\rm{fuse}}\left(\mathbf{h}^{ec}_t,  \mathbf{h}^{el}_t\right) \right) \\
	\label{eq:model:all:3}
	\mathbf{h}^{elc}_t &= \mathcal{F}_{\rm{recurrent}} \left(\mathbf{I}^{elc}_t, \mathbf{h}^{elc}_{t-1}\right) \\
	\label{eq:model:all:4}
	\mathbf{H}^{elc}_u &= \prod_{t=0}^{T} \mathbf{h}^{elc}_t
	\end{align}
	\normalsize
\end{subequations}
where multi-source sequences ($\mathcal{X}^e_u, \mathcal{X}^l_u, \mathcal{X}^c_u$) initialize the input $\mathbf{I}^*$, 
after which three levels of feature extraction and hierarchical fusion (\equationref{eq:model:all:1}, \equationref{eq:model:all:2}, \equationref{eq:model:all:3}) are conducted.
$\prod$ denotes the pooling operator that aggregates the fused representation $\mathbf{h}^{elc}$ in each temporal point, 
and the final output is the behavior embedding $\mathbf{H}^{elc}_u$ of each user $u$. 
In our experiment, we use mean pooling as the pooling operator. 
As for estimating the probability of each user being an electricity thief, 
we define a mapping function $\Psi$ that maps the feature embedding into a binary vector, 
and turn it into the probability ranging from 0 to 1 by means of the softmax function, as follows:
\begin{equation}
\mathbf{P}\left(~\mathcal{Y}~ |~ \mathcal{X}^e_u, \mathcal{X}^l_a, \mathcal{X}^c_a\right) = \rm{softmax}\left(\Psi \left(\mathbf{H}^{elc}_u\right)\right)
\label{eq:prob}
\end{equation}
We can implement $\Psi$ by means of fully connected networks or some well-known classifiers. 

The whole hierarchical framework can also be implemented by HBRNN \citep{du2015hierarchical}, which is proposed for recognizing skeleton-based actions by combining the multi-source time series data.  The main difference compared with our approach is that HBRNN implements fusion functions by directly concatenating two representations at the same temporal point; by contrast, we design a multi-step fusion mechanism for \methodshort.
We will validate the effectiveness of such architecture by comparing both methods in our experiments (\secref{sec:exp:compare}) and additional ablation studies (\secref{sec:exp:effect}).

\vpara{Learning.} 
We use the Adam optimizer~\cite{kingma2015adam} for the parameter learning, 
where the objective function is defined as the binary cross-entropy:
\begin{equation}
\begin{split}
\mathcal{L}= -\sum_{u \in \mathbf{U}} \hat{\mathcal{Y}_u}&\log\mathbf{P}\left(\mathcal{Y} |~ \mathcal{X}^e_u, \mathcal{X}^l_a, \mathcal{X}^c_a\right)
+  \\
(1 - \hat{\mathcal{Y}_u})&\log \left(1 -\mathbf{P}\left(\mathcal{Y} |~ \mathcal{X}^e_u, \mathcal{X}^l_a, \mathcal{X}^c_a\right) \right)
\end{split}
\end{equation}
where $\hat{\mathcal{Y}_u} \in \{0, 1\}$ is the ground truth of a user being an electricity thief, as confirmed by on-site investigations,
and $\mathbf{P}\left(\mathcal{Y} |~ \mathcal{X}^e_u, \mathcal{X}^l_a, \mathcal{X}^c_a\right)$ is computed by \equationref{eq:prob}.

\section{Experiments}
\label{sec:exp}
In this section, we conduct experiments on a real-world dataset (introduced in~\ref{sec:observe:data}) to answer the following three questions:
\begin{itemize} [leftmargin=*]
	\item{\textbf{Q1:}} How does \methodshort perform on electricity-theft detection tasks, compared with state-of-the-art baselines? 
	\item{\textbf{Q2:}} How does the multi-source information contribute to the detection task?
	\item{\textbf{Q3:}} Can the proposed hierarchical fusion mechanism effectively bridge the information from different sources? 
\end{itemize}

\subsection{Experimental Setup}
\label{sec:exp:setup}
\vpara{Baselines.} 
We validate the effectiveness of \methodshort~compared with several different types of baselines. 
The first type is classification methods based on handcrafted features, which have been  commonly used in existing work on electricity theft detection~\citep{xiaomei2003brief, hanmei2004analysis, guo2010survey, depuru2011electricity, viegas2017solutions}. 
We list the handcrafted features we consider in \tableref{tb:exp:feature} and employ the following classifiers: \textit{logistic regression (LR)} \citep{gortmaker1994applied}, \textit{support vector machine (SVM)} \citep{zheng2011a}, \textit{random forest (RF)} \citep{genuer2015random} and \textit{extreme gradient boosting (XGB)} \citep{chen2016xgboost}. 

The second type of baseline is time series classification methods, including: 
\begin{itemize} [leftmargin=*]
	\item \textit{Nearest Neighbor}: This method determines whether a user $u$ pilfers electrical power with reference to other users close to $u$. 
	In particular, we consider the following different metrics to calculate the distance between two users' time series in our experiment: \textit{Euclidean Distance (NN-ED)}, \textit{Dynamic Time Warping (NN-DTW)}~\citep{berndt1994using} and \textit{Complexity Invariant Distance (NN-CID)}~\citep{batista2011complexity}.
	\item \textit{Fast Shapelets (FS)} \citep{rakthanmanon2013fast}: This approach extracts shapelets, the representative segments of time series, as features for classification.
	\item \textit{Time Series Forest (TSF)} \citep{deng2013time}: This is a tree-ensemble method for time series classification. 
\end{itemize}

As for the third type of baseline, we consider the following competitive deep learning methodologies: 
\begin{itemize} [leftmargin=*]
	\item  \textit{MRNN} \citep{graves2013speech}: A multiscale recurrent neural network that takes the concatenated multi-source sequences (\small$\mathcal{X}^e_u \oplus \mathcal{X}^l_a \oplus \mathcal{X}^c_a$\normalsize) as input. 
	\item \textit{HBRNN} \citep{du2015hierarchical}: This is a hierarchical recurrent neural network on multi-source sequences, which is proposed to recognize skeleton-based actions. 
	For the fusion layer, it concatenates the latent representation at the same time points (\small$\mathcal{F}_{\rm{fuse}}(\mathbf{h}_t, \mathbf{h}'_t)=\mathbf{h}_t \oplus \mathbf{h}'_t$\normalsize).
	\item \textit{WDCNN} \citep{zheng2017wide}: A \textit{wide\&deep convolutional neural network} for detecting electricity theft that focuses on capturing periodic patterns of users' electricity usage.
	\item  \textit{\methodshort}: The proposed method. 
	We empirically set the dimension of $\mathbf{h}^e$, $\mathbf{h}^l$ and $\mathbf{h}^c$ in the first layer as 32, 8 and 8 respectively, and further set the learning rate as 0.01 with the reduction via a factor of 10 at every 20 iterations. 
	We implement $\mathcal{F}_{\rm{recurrent}}$ by LSTM, and will study how different implementations influence the performance later in \secref{sec:exp:effect}. 
\end{itemize} 

\begin{table}[t]
	\centering
	\renewcommand\arraystretch{1.}
	\addtolength{\tabcolsep}{-0pt}
	\caption{\small List of handcrafted features related to electricity theft.\normalsize}
	\begin{minipage}{0.48\textwidth}
		\small
		\begin{tabular}{m{0.2\textwidth}m{0.7\textwidth}}
			\toprule
			Feature  & Description \\
			\midrule
			
			Power usage & Mean, variance and slope of total, on-peak and off-peak electricity usage $\mathcal{X}_u^e$ \\
			NTL & Mean, variance and slope of non-technical loss $\mathcal{X}_a^l$ \\
			Temperature & Mean, variance and slope of maximum and minimum temperature $\mathcal{X}_a^c$ \\
			\midrule
			
			\multirow{2}{0.2\textwidth}{Usage vs. NTL} & $\sqrt[2]{(\mathcal{X}_u^e - \mathcal{X}_a^l)^2}$, Euclidean distance between total usage and NTL \\
			~ & $DTW(\mathcal{X}_u^e - \mathcal{X}_a^l)$, DTW distance between total usage and NTL \\
			\midrule
			
			\multirow{2}{0.2\textwidth}{Usage vs. Temperature} & $\sqrt[2]{(\mathcal{X}_u^e - \mathcal{X}_a^c)^2}$, Euclidean distance between total usage and temperatures \\
			~ & $DTW(\mathcal{X}_u^e - \mathcal{X}_a^c)$, DTW distance between total usage and temperatures \\
			
			\bottomrule 
		\end{tabular}
		\normalsize
		\label{tb:exp:feature}
	\end{minipage}
\end{table}

\vpara{Comparison metrics.}
We use precision, recall and two F-measures (F1, F0.5) as metrics. 
The F-measure is a measure of a test's accuracy and is defined as the weighted harmonic mean of the precision and recall of the test, with the following mathematical form:
\begin{equation*}
F_{\beta} = (1 + \beta^2)\frac{precision \times recall}{\beta^2 \times precision + recall}
\end{equation*}
We prefer to use F0.5 as the metric for electricity-theft detection, as precision is more important than recall to real application.

\vpara{Implementation details.}
To meet the demands of the application scenario, the input of all methods is the historical observation sequence spanning six months. The output is the probability of electricity theft, which can be validated in the next month. 
The first 80\% of samples ordered by time are used for training, and we test different methods on the remaining samples. 
We also use 10\% of samples from the training set as validation set, for avoiding the overfitting. 
For baselines that require a classifier, we use XGB \citep{chen2016xgboost} with a batch size of 2000. We adopt a larger weight (e.g., the ratio of negative/positive) for positive samples to address class imbalance.
All the experiments are ran on a single Nvidia GTX 1080Ti GPU.

\subsection{Performance Comparison}
\label{sec:exp:compare}
We first compare the experimental results of \methodshort with other baselines to answer \textbf{Q1}. 
As shown in \tableref{tb:exp:classresult}, all handcraft-feature methods perform poorly, as these methods can only capture a limited number of patterns.
Relatively speaking, the ensemble methods (i.e., \textit{XGB} and \textit{RF}) performs better (an average +7\% of F0.5). 
By automatically capturing temporal features, time series classification methods achieve further performance improvements, especially for recall. 
However, these methods cannot effectively handle multi-source time series and therefore suffer in terms of precision. 
A similar phenomenon can be observed the neural network results. In particular, when simply concatenating all multi-source data and inputting it into a recurrent neural network (\textit{MRNN}), we can see that it identifies all samples as instances of electricity theft. 
This suggests that improperly handling multi-source data will bring in more noise which hurts  performance.  
\textit{WDCNN} tries to capture the abnormal non-periodic behaviors of users, resulting in a performance improvement. 
However, the performance of this method is unstable (around $1.31$ variation).  
As expected, moreover, models with hierarchical structure like HBRNN and \methodshort are better able to handle multi-source data and consequently outperform other methods. Moreover, \methodshort outperforms HBRNN by +3\% in terms of F0.5. 
Through careful investigation, we find that with the help of the multi-step fusion and the attention operator, \methodshort can not only better bridge the multi-source information, but have superior interpretability, the details of which are presented in later chapters.

\begin{table}[t]
	\centering
	\renewcommand\arraystretch{1.}
	\addtolength{\tabcolsep}{-2pt}
	\begin{adjustwidth}{0.cm}{}
		\caption{\small Comparison of classification performance (\%). The \textbf{bold} indicates the best performance of all the methods. \normalsize}
		\label{tb:exp:classresult}
		\small
		\begin{tabular}{l|l|cccc|c}
			\toprule
			\multicolumn{2}{c|}{\textbf{\diagbox{Methods}{Metrics}}} & Precision & Recall & F1& F0.5 & Variation \\
			
			\midrule
			
			\multirow{4}{1.7cm}{Handcrafted features} & LR & 9.52 & 7.12 & 8.14  & 8.92 & $\pm$0.08  \\
			~ & SVM & 11.11 & 5.21  & 7.09 & 9.06 & $\pm$0.14  \\
			~ & RF & 17.07 & 11.93 & 14.04 & 15.72 & $\pm$0.22  \\
			~ & XGB & 20.00 & 9.35 & 12.74 & 16.29 & $\pm$0.23  \\
			
			\midrule
			\multirow{6}{1.7cm}{Time Series Classification} & NN-ED & 0.00 & 0.00 & 0.00 & 0.00 & $\pm$0.00  \\
			~ & NN-DTW & 10.82 & 27.78 & 15.57 & 12.32 & $\pm$0.12  \\
			~ & NN-CID & 12.86 & 22.96 & 16.48& 14.10 & $\pm$0.24  \\
			~ & FS & 2.26  & 20.11 & 4.06 & 2.75 & $\pm$0.85  \\
			~ & TSF & 18.67 & 24.12 & 21.05 & 19.55 & $\pm$0.53  \\
			
			\midrule
			\multirow{3}{1.7cm}{Neural Networks} & MRNN & 1.28 &  \textbf{100.0} & 2.54 & 1.59& $\pm$0.00  \\
			~ & HBRNN &  19.07 & 36.98 & 25.16 & 21.12& $\pm$0.85  \\
			~ & WDCNN & 18.46 & 24.69 & 21.12 & 19.44& $\pm$1.31  \\
			
			\midrule
			Ours & \methodshort&  \textbf{22.54} & 34.19 &  \textbf{27.17} &  \textbf{24.19} & $\pm$0.85  \\
			
			\bottomrule 
		\end{tabular}
		\normalsize
	\end{adjustwidth}
\end{table}

\subsection{Model Effectiveness}
\label{sec:exp:effect}

\vpara{Effectiveness of multi-source information (Q2).} 
We study whether or not the multi-source information can be effective in electricity theft detection. To do this, we remove the input sequences of temperature and NTL respectively from \methodshort. It is notable that after each sequence is removed, the number of \methodshort's levels decreases; once we remove both sequences, \methodshort~is transformed into a single recurrent neural network with the users' electricity usage records as input. 

From \tableref{tb:exp:multi}, we can see that multi-source information is significant: the performance drops substantially when both NTL and temperature are simultaneously removed (-12.88\% of F0.5). 
Notably, temperature is slightly less sensitive than NTL to the performance (+1.97\% of F0.5). 

\vpara{Effectiveness of hierarchical fusion mechanism (Q3).} 
The proposed hierarchical fusion mechanism includes the multi-step fusion operator and the attention operator. 
How do these elements contribute to bridging the information from different data sources?
To answer question, we remove each operator in turn and assess the subsequent impact on performance. 
Note that after removing the fusion operator, we adopt a simple concatenation for the fusion layer: $\mathcal{F}_{\rm{fuse}}=\mathbf{h}_t \oplus \mathbf{h}'_t$. 

Results are presented in \tableref{tb:exp:fusion}.  From the table, we can see that the performance clearly drops when either the multi-step fusion operator or the attention operator is removed (-2.62\% of F0.5 on average). 
Moreover, removing them both influences the performance more significantly, suggesting that these two operators work well together to bridge the information from different sources. 
We will later qualitatively demonstrate the effectiveness of our proposed hierarchical fusion mechanism through a specific application case. 

\vpara{Implementations of recurrent layers.}
Finally, we study how different implementations of the recurrent layers in our proposed model influence the performance. 
To do this, we use several common methods, such as average pooling, linear RNN~\citep{mikolov2010recurrent}, GRU~\citep{chung2015gated} and  LSTM\citep{hochreiter1997long}.
As shown in \tableref{tb:exp:rnn}, the gated networks (GRU, LSTM) outperform the linear methods (pooling, RNN), which illustrates that the appropriate temporal modeling is important for improving the performance. 

\begin{table}[tb]
	\centering
	\renewcommand\arraystretch{1.}
	\addtolength{\tabcolsep}{-0pt}
	\caption{\small Effect of multi-source information (\%).\normalsize}
	\small
	\begin{tabular}{l|cccc}
		\toprule
		Removed component & Precision & Recall & F1& F0.5 \\
		
		\midrule
		
		Temperature & 20.65 & 28.58 & 23.98  & 21.86  \\
		NTL & 19.01 & 22.33  & 20.54 & 19.59 \\
		Both & 10.46 & 16.79 & 12.89 & 11.31 \\
		
		\midrule
		\textbf{\methodshort} & \textbf{22.54} &  \textbf{34.19} &  \textbf{27.17} &  \textbf{24.19}\\
		
		\bottomrule 
	\end{tabular}
	\normalsize
	\label{tb:exp:multi}
\end{table}
\begin{table}[tb]
	\centering
	\renewcommand\arraystretch{1.}
	\addtolength{\tabcolsep}{-0pt}
	\caption{\small Effect of fusion and attention operators (\%). \normalsize}
	\small
	\begin{tabular}{l|cccc}
		\toprule
		Removed component & Precision & Recall & F1& F0.5 \\
		
		\midrule
		
		Multi-step fusion & 19.18 & 32.05  & 23.99 & 20.85  \\
		Attention & 20.91 & 30.27 & 24.73 & 22.29 \\
		Both & 18.66 & 29.47 & 22.85 & 20.14 \\
		
		\midrule
		\textbf{\methodshort} & \textbf{22.54} &  \textbf{34.19} &  \textbf{27.17} &  \textbf{24.19}\\
		
		\bottomrule 
	\end{tabular}
	\normalsize
	\label{tb:exp:fusion}
\end{table}
\begin{table}[tb]
	\centering
	\renewcommand\arraystretch{1.}
	\addtolength{\tabcolsep}{0pt}
	\caption{\small Effect of temporal modeling (\%). \normalsize}
	\small
	\begin{tabular}{l|cccc}
		\toprule
		Implementation~\qquad\qquad & Precision & Recall & F1& F0.5 \\
		
		\midrule
		
		Average pooling & 15.96 & 32.11  & 21.32 & 17.75  \\
		Linear RNN & 19.53 & 30.24 & 23.73 & 21.02 \\
		GRU & 21.04 & \textbf{36.79} & 26.77 & 23.01 \\
		LSTM & \textbf{22.54} &  34.19 &  \textbf{27.17} &  \textbf{24.19}\\
		
		\bottomrule 
	\end{tabular}
	\normalsize
	\label{tb:exp:rnn}
\end{table}

\subsection{Application and Case Study}
\label{sec:exp:case}
In the past, the State Grid staff would employ several data-driven models to detect electricity theft. However, these models are inefficient in practice.  
For example, in 2018, \textit{none} of the electricity thieves in Zhejiang were caught by these models. 
However, in order to catch thieves,  large-scale on-site investigations are very costly and time-consuming.

Accordingly, in order to improve the accuracy of on-site investigation and validate the effectiveness of our model, we employed \methodshort for monthly on-site investigation, in cooperation with State Grid Hangzhou Power Supply Co. Ltd..
More specifically, \methodshort detected 20 high-risk users at the beginning of August 2019 and suggested that the State Grid staff should investigate and collect evidence.
It turned out that our approach successfully caught three electricity thieves out of 20 with 15\% precision in practice, which represented a significant improvement in on-site investigation precision (improved from 0\%).
Moreover, another six users among the remaining 17 were identified that staff in Hangzhou strongly suspected of being electricity thieves, although no clear evidence was found during investigation.
It is likely that these six users pilfered electrical power in June or July, then restored electrical meter to its previous condition before on-site checking. 

\begin{figure}[tb]
	\begin{minipage}{0.48\textwidth}
		\centering
		\includegraphics[width=1.0\textwidth]{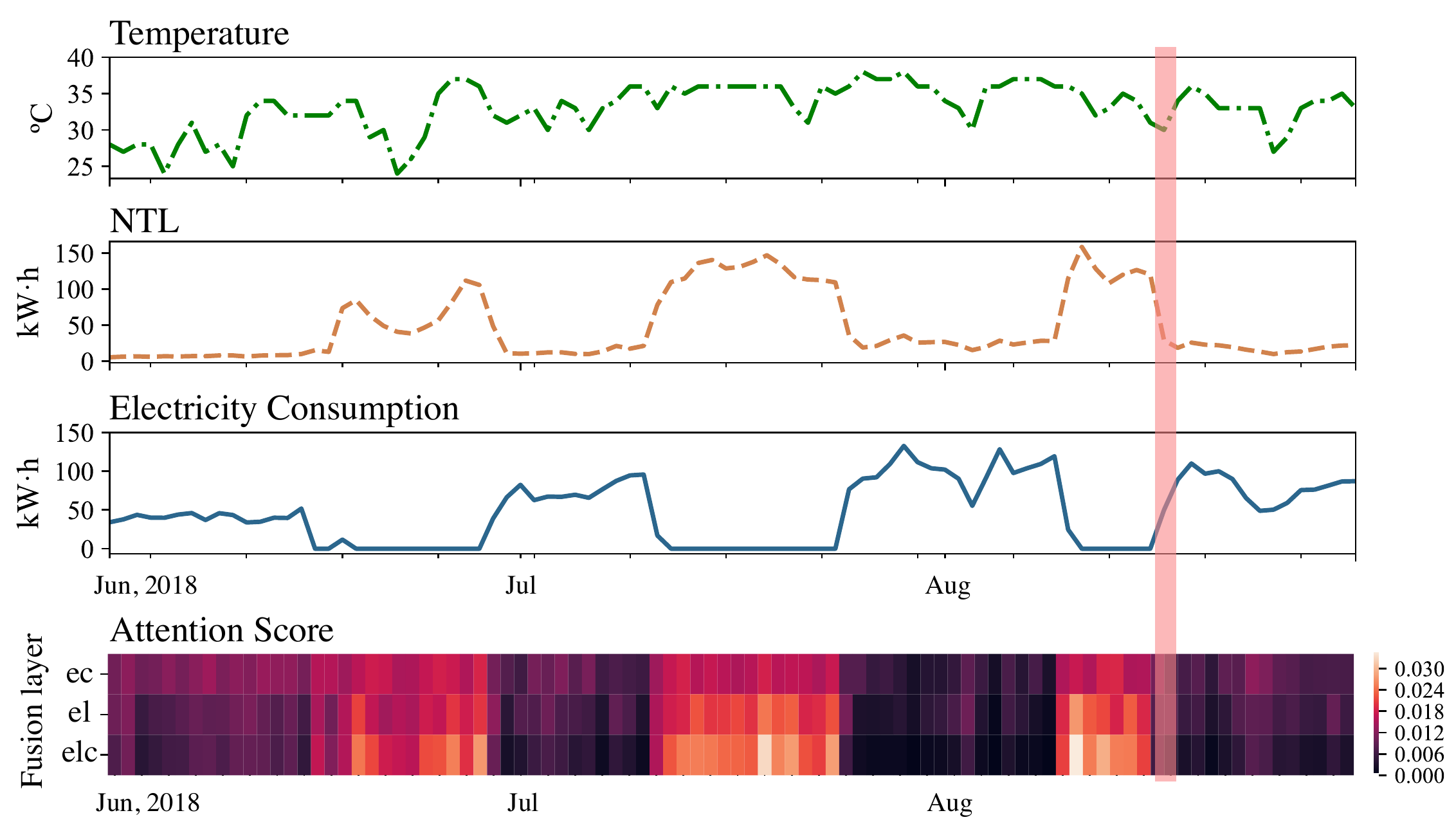}
	\end{minipage}
	\caption{\small  A case study of the attention operator shown together with different fusion layers ($\bm{\alpha}^{ec, el, elc}$). The upper rows present different multi-source observation sequences, while the red bar indicates the time at which the electricity theft was detected. The lower row presents the score vector of the attention operator by heat map. \normalsize}
	\label{fig:exp:case}
\end{figure}

We next present a specific case identified by \methodshort to demonstrate its effectiveness in real-world applications. 
In \figref{fig:exp:case}, for a specific user, we present temperature, NTL, the user's electrical usage record, and three score vectors ($\bm{\alpha}^{ec}, \bm{\alpha}^{el}, \bm{\alpha}^{elc}$) in our model's fusion layers from top to bottom.
We can see that the high-scoring positions (the brighter areas of the heat map) are where the electricity usages are low, along with hot weather and high NTL. 
This discovery is consistent with previous empirical observation (\secref{sec:observe}): when a user consumes little electricity in hot weather, while the NTL increases abnormally, he or she is likely to be pilfering electricity. 
Moreover, when examining the fusion layers from top to bottom, the brighter areas in the heat map (\figref{fig:exp:case}) become increasingly clear, suggesting that \methodshort captures more accurate patterns of multi-source information.
We can see that users do not pilfer electrical power all the time, and that the attention score returns to a low point after on-site checking (red bar).
From the case study and the performance comparison in \tableref{tb:exp:classresult}, we can see that the attention operator can not only improve the electricity-theft detection performance, but also provides superior interpretability.

\section{related work}
\label{sec:related}
\vpara{Additional related works on electricity theft.}
Electricity theft, or pilfering, is known as a common phenomenon in developing countries. Substantial effort has been expended to prevent or detect the behavior of electricity theft.
As mentioned in \secref{sec:intro}, there are two main avenues of works related to electricity theft detection or prevention.
The first of these is hardware-driven methods, and several representative works will be discussed herein.
\citet{fennell1983pilfer} proposed a special, pilfer-proofing, system incorporating a plug-in terminal block set and a meter box cover;
\citet{depuru2011electricity} first conclude that electricity theft makes up a significant proportion of non-technical loss (NTL).
Secondly, rather than attempting to improve the hardware of electricity meters,
data-driven methodologies have also been proposed that focus on the analysis of electrical power consumption records.
In addition to the works referenced in \secref{sec:intro}, 
\citet{zheng2017wide} propose a framework based on \textit{wide \& deep CNNs},
which aims at accurately identifying the periodicity and non-periodicity of electricity usage by utilizing 2-D electrical power consumption data.
Moreover, \citet{costa2013fraud} apply the use of knowledge-discovery in the database process based on artificial neural networks to conduct electricity-theft detection.
However, the problem of how to utilize multi-source data (including electricity consumption records and other related information) to conduct electricity-theft detection remains unstudied.

\vpara{Time series modeling.}
A basic but rather important characteristic of power consumption data is that these records are time series data.
Time series modeling has been widely studied over the past decades.
One traditional avenue here is to extract efficient features from the original data and develop a well-trained classifier, such as TSF~\citep{deng2013time}, \textit{shapelets}~\citep{ye2011time,lines2012shapelet} etc.;
another avenue has focused on deep learning, such as RNNs and their variants (LSTM \citep{hochreiter1997long}, GRU \citep{graves2013speech}, etc.). In addition, since the dimensions of time series features may be large, and it is likely that different levels of correlations  exist between features, hierarchical LSTM-based models are proposed to learn these hierarchical relationships (e.g., HBRNN \citep{du2015hierarchical}). 

\vpara{User behavior modeling.}
Another related domain of this work is that of the user behavior analysis.
One typical case is that  of the web search:
for example, \citet{radinsky2012modeling} develop a temporal modeling framework to predict user behavior using smoothing and trends.
However, although these works may provide various insights relevant to human behavior modeling, fully understanding complicated user behaviors is quite difficult;
thus, it may be necessary to tailor the specific analysis to the scenario in question.

\section{Conclusion and Future work}
\label{sec:conclude}
In this paper, we study the problem of electricity-theft detection and analyze the influence of the macro-level (climate) and meso-level (non-technical loss) factors on users' electricity usage behavior. 
We proposed a hierarchical framework, \methodshort, that encodes the correlations between different levels of information step by step.
When evaluated on real-world datasets, the proposed method not only achieved significantly better results than other baselines, but also helped to catch electricity thieves in practice.
In future work, we plan to explore the following aspects: 1) investigating more factors from different sources to improve electricity-theft detection accuracy; 2) extending \methodshort to other similar scenarios.

\small
\vpara{Acknowledgments.}
The work is supported by NSFC (61702447), the Fundamental Research Funds for the Central Universities, and a research funding from the State Grid of China.
\normalsize

\pagebreak
\balance
\bibliographystyle{ACM-Reference-Format}
\bibliography{main}

\end{document}